\documentstyle[prb,aps,epsf]{revtex}

\begin{document}
\title{Comparison of coherent and weakly incoherent
transport models for the interlayer magnetoresistance
of layered Fermi liquids}
\draft
\author{Perez Moses and Ross H. McKenzie\cite{email}}
\address{School of Physics, University of New
South Wales, Sydney 2052, Australia}
\date{\today}
\maketitle
\widetext
\begin{abstract}
The interlayer magnetoresistance of layered
metals in a tilted magnetic field is calculated
for two distinct models for the interlayer transport.
The first model involves coherent
interlayer transport and makes use of results
of semi-classical or Bloch-Boltzmann transport
theory. The second model involves weakly incoherent
interlayer transport where the electron is scattered many
times within a layer before tunneling into
the next layer.
The results are relevant to the interpretation of
experiments on angular-dependent magnetoresistance
oscillations (AMRO) in quasi-one- and quasi-two-dimensional organic metals.
We find that the dependence of the magnetoresistance
on the direction of the magnetic field is identical
for both models except when the field is almost parallel
to the layers.
An important implication of this result is that a
three-dimensional Fermi surface is not necessary for the observation
of the Yamaji and Danner oscillations seen in quasi-two-
and quasi-one-dimensional metals, respectively.
A universal expression is given for the dependence of
the resistance at AMRO maxima and minima on the
magnetic field and scattering time (and thus the temperature).
We point out three distinctive features of
coherent interlayer transport:
(i) a beat frequency in the magnetic oscillations of
quasi-two-dimensional systems,
(ii) a peak in the angular-dependent magnetoresistance
when the field is sufficiently large and parallel to the layers,
and (iii) a crossover from a linear to a quadratic field
dependence for the magnetoresistance when
the field is parallel to the layers.
Properties (i) and (ii) are compared with
published experimental data for a range
of quasi-two-dimensional organic metals.\\
\\
Physical Review B {\bf 60}, 7988 (1999).
\\
\end{abstract}


\vskip1pc

\section{Introduction}

One of the most fundamental concepts in
electronic transport theory for elemental
metals and semiconductors is that
electronic transport  involves the
coherent motion of electrons in band or Bloch states
associated with well-defined wave vectors
and group velocities\cite{ashcroft}.
An important issue is 
whether this concept is applicable to interlayer
transport in strongly correlated
electron systems such as
high-$T_c$ superconductors\cite{anderson,legg,hussey2},
organic conductors\cite{str}, and layered manganite
compounds with colossal magnetoresistance\cite{kimura}.
If the interlayer transport is incoherent
the motion between
layers is diffusive and it is not possible to define
band states extending over many layers
and a Fermi velocity perpendicular to the layers.                      
In that case a three-dimensional Fermi surface cannot
be defined and
Bloch-Boltzmann transport theory cannot describe the interlayer transport.

Extensive experimental studies have been made of the
angular-dependent magnetoresistance oscillations (AMRO)\cite{wos}
which occur in layered organic conductors\cite{ish} 
when the direction of the magnetic field is varied.                      
The theoretical interpretation of these oscillations often
involves a three-dimensional Fermi surface and their
observation is sometimes interpreted as evidence
for the existence of a three-dimensional Fermi surface.
In quasi-one-dimensional metals these effects are
are known as Danner\cite{dan}, Lebed (or magic angle)\cite{dan2,dan3,lebed0},
and third angular effects\cite{third},
depending on whether the magnetic field is rotated
in the ${\bf a}-{\bf c}$, ${\bf b}-{\bf c}$, or ${\bf a}-{\bf b}$
plane, respectively. 
(The most- and least-conducting directions
are the ${\bf a}$ and ${\bf c}$ axes, respectively).
In quasi-two-dimensional systems, the effects observe
include the Yamaji\cite{yam} oscillations
and the anomalous AMRO in the low-temperature phase of
$\alpha$-(BEDT-TTF)$_2$MHg(SCN)$_4$[M=K,Rb,Tl]\cite{wos,khg}.

The explaination of the Lebed effect is controversial
and a number of different theories have been proposed.\cite{str,lebed0,leb}
It is not clear that
coherent transport models can explain the angle-dependent magnetoresistance
in the quasi-one-dimensional (TMTSF)$_2$PF$_6$ at pressures of about 10
kbar\cite{str,dan2,dan3,yakov,naughton}
or the anomalous AMRO.\cite{mck}
Consequently, we focus on the Danner and Yamaji oscillations here
because their explaination in terms of
a three-dimensional Fermi surface has generally been accepted.
The resistance perpendicular to the layers is a maximum when the
field direction is such that the electron velocity
(perpendicular to the layers) averaged
over its trajectories on the three-dimensional
Fermi surface is zero\cite{dan,kart}.

Several different models for incoherent interlayer
transport have been considered previously.
We shall distinguish between what we shall refer
to as {\it weakly} and {\it strongly} incoherent
interlayer transport.
The former occurs when there is direct transfer of the
electron from one layer to another and the intralayer
momentum is conserved in the process.
Consequently, interference between the wavefunctions
on adjacent layers is possible.
However, the transport can be incoherent
in the sense that tunneling events are uncorrelated
because the electron is scattered many
times within the layer between tunneling events.
This model has been used to describe interlayer
transport in the cuprates\cite{kumar0,kumar,radtke} and
organics.\cite{soda}
In contrast, strongly incoherent transport
occurs if the the intralayer
momentum is not conserved by tunneling and
there is no interference between the wavefunctions
on adjacent layers.
This can occur because the tunneling is associated
with elastic scattering\cite{graf},
inelastic processes such as coupling to a bath
of phonons\cite{radtke} or because of non-Fermi liquid effects
such as spin-charge separation.\cite{str,kumar1}
For both weakly and strongly incoherent
transport, the interlayer conductivity of a bulk sample
is determined by the tunneling rate between two adjacent
layers. The interlayer resistance is then equal to the
number of layers in the sample times the resistance
between two layers.

In this paper we present the details of calculations
of the angular dependence of the interlayer magnetoresistance
for both coherent and weakly incoherent interlayer transport
when there is a Fermi liquid within each layer.
Our main result is that coherent interlayer transport is not {\it
necessary}
to explain the Yamaji and Danner oscillations.
Hence, their observation is {\it not} evidence for the existence
of a three-dimensional Fermi surface.
In contrast, we point out three properties of the
interlayer magnetoresistance which do only occur if
the interlayer transport is coherent:
(i) a beat frequency in the magnetic oscillations
of quasi-two-dimensional systems,
(ii) a peak in the angle-dependent magnetoresistance
when the field is parallel to the layers for sufficiently high fields,
and (iii) a crossover from linear to quadratic field dependence
when the field is parallel to the layers.
A brief report of some of the results presented here appeared
previously.\cite{mckenzie}

In the next section we present our main result, Eq.~(\ref{result}), an
analytical
expression for the interlayer conductivity in the presence of
a magnetic field which is tilted at an angle $\theta$ relative
to the normal to the layers.
This result is valid for incoherent transport for
all field directions and for coherent transport
provided the field is not almost parallel to the
layers. We then use this expression to explain the
basic features of the Danner and Yamaji oscillations.
Simple expressions are described for the dependence of
the interlayer resistance on the magnitude of the
magnetic field and the scattering rate when the angle $\theta$
is at an AMRO maxima or minima.
In Section \ref{coherent} we derive Eq.~(\ref{result}) for the case of
coherent interlayer transport for both quasi-two- and
quasi-one-dimensional systems. This involves evaluating
Chambers' formula, a result of Bloch-Boltzmann transport theory.
In Section \ref{incoherent}
we derive Eq.~(\ref{result}) for weakly incoherent
transport in both quasi-two- and quasi-one-dimensional systems.
In Section \ref{sig} we consider unambiguous signatures
of coherent interlayer transport and compare these
signatures to published experimental data for a range
of quasi-two-dimensional metals.

\section{ANGULAR DEPENDENT MAGNETORESISTANCE OSCILLATIONS}

We assume that each layer of the metal
is a Fermi liquid whose elementary excitations are
fermions with wave vector $(k_x,k_y)$ and
with a dispersion relation of the
form $\epsilon(k_x,k_y)$. We consider the simplest
posssible dispersion relations for quasi-one- and
quasi-two-dimensional systems.
(For a summary of our notation
see Table I  in Ref. \onlinecite{mckenzie}).
The interlayer conductivity in the
absence of a magnetic field
is denoted $\sigma_{zz}^0$.
We will show in this paper that in a tilted magnetic field
the interlayer conductivity, for both
coherent interlayer transport (except
for fields very close to the layers)
and weakly incoherent interlayer transport
(for all field directions), is
\begin{equation}
\sigma_{z z} (\theta) =\sigma_{z z}^{0} \hspace{1mm}
\left[ J_{0}(\gamma \tan\theta)^{2} +
2 \sum_{\nu=1}^{\infty}
{ J_{\nu}(\gamma \tan\theta)^{2} \over
{1 + (\nu \omega_0 \tau \cos \theta )^2} } \right]\ ,
\label{result}
\end{equation}
where $J_{\nu}(x)$ is the $\nu$-th order
Bessel function, $\omega_0$ is the oscillation frequency
associated with the magnetic field, and
$\gamma $ is a constant that depends on the geometry of
the Fermi surface.\cite{mckenzie}     
The scattering time $\tau$ is assumed to be independent
of the momentum of the electron but can vary with temperature.

If the field is sufficiently large and the
temperature sufficiently low that
$\omega_0 \tau \gg 1$ then the first term in (\ref{result})
is dominant.
However, if $\gamma \tan \theta$ equals a zero of
the zero-th order Bessel function then
at that angle $\sigma_{zz}$ will be a minimum
and the interlayer resistivity will be a maximum.
If $\gamma \tan \theta \gg 1$, then the zeroes occur
at angles $\theta_n$ given by
\begin{equation}
\gamma \tan \theta_n = \pi (n - {1 \over 4}) \ \ \ \ \ (n=1,2,3, \cdots)\ .
\label{minangle}
\end{equation}
This condition was first derived for the
quasi-two-dimensional case by Yamaji\cite{yam} and for
the quasi-one-dimensional case by Danner, Kang, and Chakin\cite{dan}.
Determination of these angles experimentally provides
a value for $\gamma$ and thus information about the intralayer Fermi
surface.
The values of the Fermi surface area of quasi-two-dimensional systems
determined from AMRO are in good agreement with the
Fermi surface areas determined from the frequency
of magneto-oscillations\cite{wos}.
Furthermore, AMRO can be used to map out the actual shape
of the Fermi surface within the layer (see for example
References \onlinecite{kart,kovalev,yak2}).

The angular dependence of the interlayer resistivity
$\rho_{zz} \simeq 1/ \sigma_{zz}$\cite{caveat},
given by Eq.~(\ref{result}),
for parameter values relevant for
typical quasi-two-dimensional systems is
shown in Fig. \ref{plots1}.
Fig. \ref{plots2} shows the angular
dependence of
$\rho_{zz}$
for parameter values relevant to
(TMTSF)$_2$ClO$_4$.
The results are similar to the experimental results
in Ref. \onlinecite{dan}, except near 90 degrees.
Both figures are very similar to the results
of numerical integration of Chambers' formula
for coherent transport (Eq.~(\ref{eq:chambers}))
except near 90 degrees. For coherent transport there is
a small peak in $\rho_{zz}(\theta)$ at $\theta = 90$
degrees.\cite{dan,hanasaki}
This is due to the existence of closed orbits on the Fermi surface when
the field lies close to the plane of the layers.
For incoherent tranport these orbits do not exist
and so the associated magnetoresistance is not present.
Since Eq.~(\ref{result}) is also valid for incoherent
interlayer transport
the Danner and Yamaji oscillations can be explained equally well in
terms of weakly incoherent transport.

\subsection{Asymptotic Form}
We want to find an expression for
$\sigma_{z z}(\theta)$ as
$\theta \rightarrow {\pi \over 2}$.
Using the asymptotic form~\cite{abram}
\begin{equation}
J_{n}(z) = \sqrt{2\over \pi z}
\cos \left( z - { n \pi \over 2} -
{\pi \over 4}\right) \label{bs1}
\end{equation}
which is valid for $z >> n^2$, we can
simplify Eq.~(\ref{result}) for the conductivity.
Re-writing it as
\begin{equation}
\sigma_{z z} =\sigma_{z z}^{0}
\hspace{1mm} \left[ J_{0}(\mu)^{2} +
2 \sum_{\nu=1}^{\infty}
{ J_{\nu}(\mu)^{2} \over {1 + (\nu x)^2}
}\right]\ , \label{bs2}
\end{equation}
where $\mu = \gamma \tan\theta$ and
$x = \omega_0 \tau \cos\theta$.
We can substitute
Eq.~(\ref{bs1}) for $J_n(\mu)$ provided
that $x > 1$ so that the sum in Eq.~(\ref{bs2})
converges rapidly. Separating the sum
into the sum of even and odd terms, and
using the fact that $2\cos^2\left( \mu - { n \pi \over 2} -
{\pi \over 4}\right)= 1+ \sin(2 \mu)$ for even $n$
and $1- \sin(2 \mu)$ for odd $n$, gives
\begin{equation}
\sigma_{z z} ={ 2\sigma_{z z}^{0}\over
\pi \mu} \hspace{1mm} \left[
{(1+ \sin 2\mu)}{\left({1\over 2} +
{\sum_{n = 1}^{\infty}}
{1\over{1+(2 n x)^2}}\right)} +
{(1- \sin 2\mu)}{\left( \sum_{n = 0}^{\infty}
{1\over{1+(2 n+1)^2 x^2}}\right)}
\right]\ . \label{bs5}
\end{equation}
These series can be evaluated using
the residue theorem~\cite{spi} to give
\begin{equation}
{\sigma_{z z} \over \sigma_{z z}^{0}} ={ 1\over
\mu x} \hspace{1mm} \left[ \coth \left({ \pi\over x}\right)
+ { \sin(2 \mu)\over \sinh({ \pi \over x}) }
\right]\ . \label{bs12}
\end{equation}
For $\omega_0 \tau > 1$, Eq.~(\ref{bs12})
actually turns out to be a good approximation
for $\gamma \tan\theta > 1$ (see Fig. 1).
It will now be used to analyse the
field and temperature dependence of the AMRO maxima and minima.

\subsection{Field and temperature dependence of the resistivity at
critical angles}

{\it Resistivity maxima:} For the
resistivity $ \rho_{zz}$
to be a maximum
$\theta = \theta^n_{max}$ where
\begin{equation}
\gamma \tan\theta^n_{max} = \left(n - { 1\over 4} \right) \pi \ .
\end{equation}
From this we can simplify $ \sin (2 \mu)$
in Eq.~(\ref{bs12}) giving
\begin{equation}
\sin(2 \mu)=
\sin(2 \gamma \tan\theta)= -\cos(2 n \pi)= -1
\end{equation}
for all $n$. The resistivity
is then written as
\begin{equation}
{\rho_{zz}(\theta^n_{max},B) \over
\rho_{zz}(B = 0) \left(n - {1 \over 4}\right)\pi} =
{ \omega_0 \tau \cos\theta_{max}
\over \tanh \left( 
{ \pi \over 2 \omega_0 \tau \cos\theta_{max}}\right)}\ .  \end{equation} This expression is plotted in Fig. \ref{p2theta}. 
Now if the field is sufficiently high and
the temperature sufficiently low that
$\omega_0 \tau \cos\theta^n_{max}>> 1$, then 
the resistivity becomes
\begin{equation}
{\rho_{zz}(\theta^n_{max}) \over \rho_{zz}(B = 0)}= 
{\gamma \over \pi }(\omega_0 \tau)^2 \sin(2 \theta^n_{max})\ .
\end{equation}
Hence, at a fixed field the resistivity at the AMRO maxima
will have the same temperature dependence as the
scattering time which is inversely proportional to 
the zero-field resistivity.

{\it Resistivity minima:} Similar arguments will show that for
the interlayer resistivity to be a minimum
$\theta = \theta^n_{min}$ where
\begin{equation}
\gamma \tan\theta^n_{min} = \left(n +{ 1\over 4} \right) \pi 
\end{equation}
and the resistivity then becomes
\begin{equation}
{\rho_{zz}(\theta^n_{min},B ) \over 
\rho_{zz}(B = 0)\left(n +{ 1\over 4} \right) \pi}= 
\omega_0 \tau \cos\theta^n_{min} 
\tanh \left( 
{ \pi \over 2 \omega_0 \tau \cos\theta^n_{min}}\right)\ .
\end{equation}
This is plotted in Fig. \ref{p2theta}. 
When $\omega_0 \tau \cos\theta_{min} >> 1$ then
\begin{equation}
{\rho_{zz}(\theta_{min}) \over 
\rho_{zz}(B = 0)}= 
{\pi^2 \over 2} \left(n +{ 1\over 4} \right) 
\end{equation}
which is independent of the field and scattering rate. 
Thus, at the AMRO minima the resistance will
have the same temperature dependence as 
the zero-field resistance.

{\it Field in the layers:}
Now as $\theta \rightarrow
{ \pi \over 2}$, $x \mu \rightarrow \gamma \omega_0 \tau$,
$x \rightarrow 0$, and
taking these limits in (\ref{bs12}) gives
\begin{equation}
\sigma_{zz} \left(\theta =
{ \pi \over 2}\right) =
{\sigma_{zz}^{0} \over \gamma \omega_0 \tau}.
\label{linear}
\end{equation}
The resistivity is
linear in field at moderately high
fields. 
However, caution is in order because in deriving (\ref{bs12})
above we required that $x > 1$. 
That this is more restrictive than need be is suggested
by the fact that Fig. 1 shows that (\ref{bs12}) remains 
valid near 90 degrees.
Indeed, Eq. (\ref{linear}) is
valid: it
agrees with the calculations of other authors
for both  coherent~\cite{scho}
and incoherent transport\cite{macdonald} in a
quasi-two-dimensional system, provided the field is not to large
(see Sections IV B and V C).
Such a linear interlayer
magnetoresistance has been observed
in Sr$_2$RuO$_4$ (Ref. \onlinecite{hussey}) and
(TMTSF)$_2$ClO$_4$ (Ref. \onlinecite{danner}).
However, for coherent transport the dependence
on field becomes quadratic
for sufficiently high fields\cite{scho} (see Section \ref{sig}).
Note that (\ref{linear}) is actually independent of the scattering time
$\tau$.
This means that for moderate fields
the interlayer resistivity will only depend weakly on temperature.
This was observed in
(TMTSF)$_2$ClO$_4$:
when the temperature was increased from 0.9 K to 8 K
the zero-field resistivity (which is proportional
to the scattering rate) increased by a factor of
about six but the resistance at 12 Tesla
increased by less than ten per cent.

\section{COHERENT INTERLAYER TRANSPORT}
\label{coherent}
If the interlayer transport is coherent
one can define a wave vector $k_z$ perpendicular
to the layers and a
three-dimensional dispersion relation 
$\epsilon_{3D}(\vec k)$
of the form 
\begin{equation} 
\epsilon_{3D}(\vec k)= \epsilon(k_x,k_y) - 2 t_c \cos (k_z c)\ , 
\label{eq:disp} 
\end{equation}
where $t_c$ is the interlayer hopping integral,
$c$ is the layer separation, and
$\epsilon(k_x,k_y)$
is the intra-layer dispersion
relation, simple examples of which are given below.
The electronic group velocity perpendicular to
the layers is
\begin{equation}
v_z = {1 \over \hbar} {\partial \epsilon(\vec k)
\over \partial k_z}
= { 2 c t_c\over \hbar}\sin(c k_z)\ .
\label{eq:groupvel}
\end{equation}
We calculate the interlayer conductivity
by solving the Boltzmann equation in the relaxation
time approximation which leads to Chambers' formula\cite{ashcroft}
\begin{equation}
\sigma_{zz} = {e^2 \tau \over 4 \pi^3}
\int { v_z(\vec k) \bar{v}_z(\vec k)}
\left(-{\partial f(E) \over
\partial E}\right) d^3\vec k \ ,
\label{eq:chambers}
\end{equation}
where $f(\epsilon)$ is the Fermi function and $\tau$
is the scattering time
which is assumed to be the same at all
points on the Fermi surface.
$\bar{v}_z(\vec k)$ is the electron
velocity perpendicular
to the layers and is averaged over
its trajectories on the Fermi
surface

\begin{equation}
\bar{v}_z(\vec k) = {1 \over \tau}\int_{-\infty}^{0}
\exp \left({t \over \tau}\right)
v_z(\vec k(t)) dt \ . \label{eq:vbar}
\end{equation}
where $\vec k(0) = \vec k$.
The time dependence of the wave
vector $\vec{k}(t)$ is found by
integrating the semi-classical
equation of motion
\begin{equation}
{d \vec k \over dt} =
-{e \over \hbar^2} \vec{\nabla}_k
\epsilon_{3D} \times \vec B \ . \label{prop}
\end{equation}
Now if the temperature is sufficiently
low that $T << E_F$ then
${\partial f \over \partial E}$ in
Eq.~(\ref{eq:chambers})
can be replaced by a delta function
at the Fermi energy and Eq.~(\ref{eq:chambers})
becomes
\begin{equation}
\sigma_{zz} = {e^2 \tau \over 4 \pi^3}
\int v_z(\vec k) \bar{v}_z(\vec k)
\delta(E_F-\epsilon_{3D}(\vec{k})) d^3\vec{k}\ .
\label{chambers2}
\end{equation}

\subsection{Quasi-two-dimensional case}

Here we consider a quasi-two-dimensional system
with the energy dispersion relation
\begin{equation}
\epsilon_{3D}(\vec k)= {\hbar^2 \over 2 m^\star}
{\left(k_x^2 + k_y^2\right) } - 2t_c
\cos \left(k_z c \right ) \ , \label{eq:dispersion}
\end{equation}
where $m^\star$ is the effective mass
of the electron.
We assume the interlayer hopping is
sufficiently small that
$t_c \ll E_F$.
The Fermi surface is then a warped
cylinder(see Ref. (\onlinecite{mckenzie})).
Substituting the energy dispersion relation from
Eq.~(\ref{eq:dispersion})
we obtain the components of the
group velocity
\begin{equation}
\vec v(\vec k) = { 1\over \hbar} \vec{\nabla}_k \epsilon_{3D}=
\left( {\hbar k_x \over m^\star} ,
{\hbar k_y \over m^\star} ,
{2 c t_c \sin(c k_z)\over \hbar} \right)\hspace{3pt}.
\label{eq:xyzcopmv}
\end{equation}
In order to calculate the time dependence
of $k_z$ we must integrate
Eq.~(\ref{prop}) which can be written in the form
\begin{equation}
{d \vec k \over dt} = {e \over m^\star}
\left( -k_y B\cos\theta,
k_x B\cos\theta, k_y B\sin\theta
\right)\ . \label{eq:wavect}
\end{equation}
Terms of order $t_c$ have been neglected
once we assume $t_c \tan\theta \ll\ {\hbar^2 k_F \over m^* c}$,
where $k_F$ is the Fermi wave vector, defined by
$E_F = {\hbar^2 k_{F}^{2} \over 2 m^* }$.
Differentating the $x$ and $y$ components of
Eq.~(\ref{eq:wavect}) with respect to time
we obtain a second order differential
equation whose solution gives
$k_y(t)=k_F \cos(\omega_c t)$ and
$k_x(t)=k_F \sin(\omega_c t)$, and

\begin{equation}
\omega_c = {e B\cos\theta \over m^\star }
\end{equation}
is the cyclotron frequency. Substitution
of this into the $z$
component of Eq.~(\ref{eq:wavect})
and integrating gives
\begin{equation}
k_z(t) = {k_z(0) + k_F \tan\theta \sin(\omega_c t)} \ .
\end{equation}
In order to calculate the $z$ component
of the group velocity we substitute
the expression for $k_z(t)$ into the
$z$ component of Eq.~(\ref{eq:xyzcopmv})
giving
\begin{equation}
v_z(k_z(0), \phi) =
C \sin(\mu \sin\phi + k_z(0) c) \ , \label{eq:vz}
\end{equation}
where
\begin{equation}
\mu = c k_F \tan\theta \ ,
\end{equation}
$\phi$ is the angle around the orbit, and $C = 2ct_c/\hbar$.
Integrating the velocity in Eq.~(\ref{eq:vz})
over a period gives us the
average velocity which can be written as
\begin{equation}
<v_z> \propto \int_{0}^{2 \pi} \sin(\mu \sin\phi)
d\phi \sim J_0(\gamma \tan\theta)
\end{equation}
and in the absence of scattering this average
velocity is equal to zero when
$\gamma \tan\theta$ equals a zero of $J_0$. These
particular values of $\theta$ corrrespond
to the peaks in the resistivity.

We can write Eq.~(\ref{eq:chambers}) in a slightly
simplified form in order to highlight the
fact that the integral is a surface
integral. If the warping of the Fermi surface
is small we can parametrise the surface
using $k_z$ and $\phi$
where $k_x=k_F \cos \phi$ and
$k_y=k_F \sin \phi$ giving
\begin{equation}
\sigma_{z z} = {e^2 \tau m^* \over
4 \pi^3 \hbar^2} \hspace{2pt} \int_{FS}
dS \hspace{2pt} v_z(\vec k) \bar{v}_z
(\vec k) ={e^2 \tau m^* \over 4 \pi^3 \hbar^2}
\hspace{1pt} \int_{-\pi/c}^{\pi/c} dk_z
\int_{0}^{2 \pi} d\phi \hspace{1pt}
v_z(\vec k) \bar{v}_z (\vec k) \;
\end{equation}
here $\bar{v}_z (\vec k)$ is defined in
Eq.~(\ref{eq:vbar})
and the pre-factor $m^* /\hbar^2$ arises from
the delta function . In terms
of the parametrised surface we have
\begin{equation}
\bar{v}_z(\vec k) = \int_{-\infty}^{0}
{d\phi' \over \tau \omega_c}
\exp({\phi' / \tau \omega_c})
\hspace{4pt} v_z(k_z(0), \phi - \phi') \ , \label{eq:vb}
\end{equation}
where we used the fact that
$\phi'= \omega_c t$ and $v_z(\vec k(t))=
v_z(k_z(0), \phi - \phi')$. For
closed electron orbits the
electron group velocities are periodic
functions of $\phi$ and $\phi'$.
Thus the range of integration of $\phi'$
can be cut up into segments each having
length 2$\pi$. The conductivity is then
\begin{eqnarray}
\sigma_{z z} ={e^2 m^* \over 4 \pi^3 \hbar^2}
\int_{-\pi/c}^{\pi/c} dk_z(0) {1\over
(1-\exp({-{2 \pi / \tau\omega_c }})}
\int_{0}^{2 \pi} \hspace{1pt} d\phi
\hspace{2pt} {v_z(k_z(0), \phi)}
\times \nonumber \\
\int_{-2 \pi}^{0} {d\phi' \over \omega_c }
\hspace{2pt} v_z( k_z(0), \phi - \phi')
\exp({\phi' / \tau\omega_c}) \hspace{3pt}. \label{eq:fincond}
\end{eqnarray}
We use trigonometric identities to
expand Eq.~(\ref{eq:vz}) and substitute
the Bessel generating
functions~\cite{abram} to obtain
\begin{eqnarray}
v_z(k_z(0), \phi-\phi') &=& C \hspace{1mm}
\sin(k_z(0) c) \hspace{1mm} \left[J_0(\mu)+
2 \sum_{k=1}^{\infty} J_{2k}(\mu) \cos((2 k)
(\phi-\phi'))\right] + \nonumber \\
& & C \hspace{1mm} \cos(k_z(0) c)
\hspace{1mm}
\left[2 \sum_{k=0}^{\infty} J_{2k+1}(\mu)
\sin((2 k+1) (\phi-\phi'))\right] \hspace{3pt}.
\end{eqnarray}
If we substitute this into Eq.~(\ref{eq:vb})
we obtain
\begin{eqnarray}
\bar{v}_z(\phi) & = & C \int_{-\infty}^{0}
{d\phi' \over \tau \omega_c} \hspace{2mm}
\left( \hspace{1mm} \sin(k_z(0) c) \hspace{1mm}
\left[J_0(\mu)+
2 \sum_{k=1}^{\infty} J_{2k}(\mu) \cos((2 k)
(\phi-\phi'))\right] \right. \nonumber \\
& & \left. + \hspace{1mm} \cos(k_z(0) c)
\hspace{1mm}
\left[2 \sum_{k=0}^{\infty} J_{2k+1}(\mu)
\sin((2k+1) (\phi-\phi'))\right] \right)
\exp(\phi^{'}/\tau \omega_c) \hspace{3pt}. \label{eq:fullvb}
\end{eqnarray}
Substituting equations for $v_z(k_z(0),
\phi-\phi')$ and $v_z(k_z(0), \phi)$
into Eq.~(\ref{eq:fincond}) we note,
terms that survive is when
$k = l$, since integrals such as
$\int_{0}^{2\pi} d\phi \cos(2k\phi) \cos(2l\phi)$ =
$\pi \delta_{kl}$, where $\delta_{kl}$
is the Kronecker delta, thus giving

\begin{eqnarray}
\sigma_{z z} & = & {e^2 (2 c t_c)^2 m^* \over
4 \pi^3 \hbar^4 \omega_c}
\int_{-\pi/c}^{\pi/c} dk_z(0) \hspace{2mm}
\left[ 2 \pi \tau \omega_c \sin(k_z(0) c)^{2}
J_{0}(\mu)^{2}
\right. \nonumber \\
& & +{4 \pi \over \omega_c \tau} \left.
\left(\sin(k_z(0) c)^{2} \sum_{k=1}^{\infty}
{J_{2k}(\mu)^{2}
\over {(2k)^2 + (1/\omega_c \tau)^2}} +
\cos(k_z(0) c)^{2} \sum_{k=0}^{\infty}
{J_{2k+1}(\mu)^{2}
\over {(2k +1)^2 + (1/\omega_c \tau)^2} }\right) \right]\hspace{3pt}.
\end{eqnarray}
Performing the integral over $k_z(0)$ yields the final
expression for the conductivity
which is of the form of Eq. (1).
This result was previously given by
Yagi {\em et al.}~\cite{yagi}.

\subsection{Quasi-one-dimensional case}

For the quasi-one dimensional case we begin
with the dispersion relation~\cite{dan}
\begin{equation}
\epsilon_{3D}(\vec k) ={\hbar v_F(|k_x| -k_F)} -
2t_b \cos(k_y b) - 2t_c \cos(k_z c) \
\end{equation}
where $v_F$ is the Fermi velocity and $t_b$
is the intrachain hopping within the layers.
The Fermi surface consists of two sheets at
$k_x \cong \pm k_F$.
By proceeding as for the
quasi-two-dimensional case the rate
of change of the wave vector (and defining
the magnetic field
$\vec B = (B\sin \theta, 0, B\cos \theta)$)
is given by

\begin{equation}
{d\vec k \over dt}= { 1\over \hbar^2}
\pmatrix{-2 b B \cos \theta e t_b \sin(b k_y) \cr
{e B \hbar \cos \theta v_F } \cr
2 b B \sin \theta e t_b \sin(b k_y)} \hspace{100pt}
\pmatrix{a \cr b \cr c} \ , \label{genk}
\end{equation}
where we neglect terms involving $t_c$
by assuming that $v_F \gg v_z \tan\theta$,
i.e., the warping of the Fermi
surface is small and the magnetic field is
not too close to the layers.
Integrating the second equation, gives
\begin{equation}
k_y(t) = { \omega_{B}\over b} t + k_y(0) \ , \label{ky}
\end{equation}
where
\begin{equation}
\omega_{B} \equiv {e B b \cos\theta v_F \over \hbar }
\label{omb}
\end{equation}
is the speed at which the wave vector traverses the
Fermi surface. To obtain $k_z(t)$ we
substitute Eq.~(\ref{ky}) into
(\ref{genk}c) and integrate to obtain
\begin{equation}
k_z(t) = k_z(0) -
{2 e b t_b B \sin\theta \over \hbar^2 \omega_B}\cos(\omega_B t +
b k_y(0))\hspace{3pt}.
\end{equation}
Substitution into the $z$-component of the
velocity yields
\begin{equation}
v_z(k_z(0), \phi-\phi^{'}) = {2 c t_c \over \hbar}
\sin( c k_z(0) - \gamma \tan\theta \cos(\phi-\phi^{'})) \ ,
\end{equation}
where $\phi^{'}= -\omega_B t$, $\phi = b k_y(0)$ and
$\gamma = { 2 c t_b \over v_F \hbar}.$
This is similar in form to the $z$ component
of the velocity for the quasi-two
dimensional case (compare with Eq.~(\ref{eq:vz})).
The interlayer conductivity can then be
written as

\begin{eqnarray}
\sigma_{z z} = {e^2 \over 4 \pi^3 b \hbar v_F }
\int_{-\pi/c}^{\pi/c} dk_z(0)
\int_{-\pi/b}^{\pi/b} d\phi \hspace{3pt}
{v_z(k_z(0), \phi)} \cr
\times \int_{-2 \pi}^{0} { d\phi^{'}\over \omega_B}
{e^{\phi^{'}/ \tau\omega_B}
\over (1-\exp({-{2 \pi / \tau\omega_B}}))}
\hspace{3pt} v_z(k_z(0), \phi-\phi^{'}) \hspace{3pt}.
\end{eqnarray}
The integral from $-2 \pi$ to 0 over $\phi^{'}$
is obtained
by noting that the electron group velocity is a
periodic function of $\phi^{'}$, we can cut
the range of $\phi^{'}$
into segments having length $2 \pi$.
The factor $(1-e^{-{2 \pi / \tau\omega_B}})$
is a consequence of this. Proceeding as for
the quasi-two dimensional case leads to
the result of the form Eq.~(\ref{result}).
As far as we are aware this expression has not been
derived previously for quasi-one-dimensional systems.

\section{WEAKLY INCOHERENT INTERLAYER TRANSPORT}
\label{incoherent}

Suppose that the coupling between the layers is sufficiently
weak that the time it takes an electron to hop between the
layers (approximately $\hbar / t_c$)
is much longer than the scattering time.
This means that the intralayer scattering rate is much larger
than the interlayer hopping integral
\begin{equation}
{\hbar \over \tau} \gg t_c
\end{equation}
and the mean-free path perpendicular to
the layers is much smaller than the interlayer
spacing. If this condition holds then
the interlayer transport will be incoherent
in the sense that successive interlayer tunneling
events are uncorrelated\cite{kumar0}. Previous estimates
of $t_c$ and $\tau$ in various layered organic
metals\cite{wos,dan,balthes} suggest
these quantities may be comparable at low
temperatures. Furthermore, the scattering rate usually
increases quadratically with temperature\cite{dressel2}
and at temperatures of the order
of tens of Kelvin this condition will almost certainly be
satisfied.\cite{moser}
The interlayer conductivity
is then proportional to the tunneling rate
between just two adjacent layers.
If we assume that the intralayer momentum
is conserved the tunneling rate can be calculated using standard
formalisms
for tunneling in metal-insulator-metal
junctions.\cite{mahan}
Modelling the interlayer transport in this way is reasonable
because many organic conductors
consist of conducting layers separated by insulating layers
of anions that are several $\AA$ thick.
Furthermore, intrinsic Josephson type
effects have been observed in the superconducting state of
$\kappa$-(BEDT-TTF)$_2$Cu(NCS)$_2$ (Ref. \onlinecite{mansky}).

We consider the simplest possible model for
the tunneling between layers, direct transfer
described by the Hamiltonian
\begin{equation}
H_{12} = - t_c \int d^2 \vec{r} \ \ \left(c_1(\vec{r})^\dagger
c_2(\vec{r}) + c_2(\vec{r})^\dagger c_1(\vec{r})\right) \ ,
\end{equation}
where $c_1(\vec{r})^\dagger$ creates an electron in layer
1 at $\vec{r}$.
Note that the interlayer transport is coherent
in the sense that during the tunneling process the
phase information in the electron's wave function is
not completely lost.
However, it is incoherent in the sense that due to
the large intralayer scattering rate
the interlayer transport cannot be described by
Bloch states extending over many layers.
If we consider a sequence of tunneling events they
are uncorrelated because after a tunneling
event an electron is scattered
many times before it tunnels to the next layer.

The interlayer current $I$ associated with $H_{12}$
and produced by a voltage $V$
can be calculated
using the formalism developed for
metal-insulator-metal junctions.\cite{mahan} The
result for the interlayer conductivity is
\begin{equation}
\sigma_{zz} = { c \over L_x L_y }\left.{ dI\over dV}
\right|_{V=0} = { 2 e^2 t_{c}^{2} c \over
\hbar L_x L_y } \int d^2r \int d^2r^{'}
\int {dE \over 2 \pi}
A_{1}(\vec{r},\vec{r^{'}},E) A_{2}(-\vec{r},\vec{r^{'}},E)
{\partial f(E) \over \partial E} \ ,
\end{equation}
where $A_{1}$ and $A_{2}$ are the
spectral functions for layers
1 and 2 and $L_x$ and $L_y$ are the
dimensions of the layers.
It will be seen below that in the presence of a tilted magnetic
field $A_1$ and $A_2$ are not identical.
The zero-field limit (for which $A_1=A_2$) of this expression
has been used in treatments of incoherent interlayer
transport in the cuprate superconductors\cite{kumar,radtke,kumar1}.

If we assume $T << E_F$, then
${\partial f(E) \over \partial E}$ can be replaced with
a delta function to give
\begin{equation}
\sigma_{zz} = {e^2 t_{c}^{2} c \over \hbar \pi L_x L_y }
\int d^2r \int d^2 r^{'} A_{1}(\vec{r},\vec{r^{'}}, E_F)
A_{2}(\vec{r^{'}},\vec{r}, E_F)\hspace{3pt}. \label{condA}
\end{equation}
This can be re-written by noting that
the spectral function can be written as
$A_{1,2}(\vec{r^{'}},\vec{r}, E_F) =
-{ 1 \over\ i } [G_{1,2}^{+}(\vec{r},
\vec{r^{'}}, E_F)-G_{1,2}^{-}(\vec{r^{'}},\vec{r}, E_F)]$
leading to
\begin{equation}
\sigma_{zz}(\vec{r^{'}},\vec{r}, E_F)
= { e^2 t_{c}^{2} c \over
\hbar \pi L_x L_y} \int d^2r \int d^2r^{'}
[G_{1}^{+}(\vec{r},\vec{r^{'}}, E_F)
G_{2}^{-}(\vec{r^{'}},\vec{r}, E_F)+
G_{1}^{-}(\vec{r^{'}},\vec{r}, E_F)
G_{2}^{+}(\vec{r},\vec{r^{'}}, E_F)] \ . \label{condG}
\end{equation}

In the Landau gauge, the vector potential $\vec{A}$, for the
magnetic field $\vec{B} = (B_x, 0, B_z)=
(B\sin\theta ,0, B\cos\theta)$, is
\begin{equation}
\vec{A} = (0, xB_z - zB_x, 0) \ ,
\end{equation}
where $\vec{B}$ and $\vec{A}$ are related by
$\vec{B}=\vec{\nabla} \times \vec{A}$. The vector
potential in the two layers (see Ref. (\onlinecite{mckenzie}))
are not equal and differ by a gauge transformation
$\vec{A}_{2}=\vec{A}_{1}+ \vec{\nabla} \Lambda$
where
\begin{equation}
\vec{\nabla} \Lambda=\vec{A}_{1}-\vec{A}_{2} =
(0, -c B\sin\theta, 0)\hspace{3pt}.
\end{equation}
The Green's functions in the two layers
are not equal.
This reflects the fact that even though
the magnetic field is invariant
under a gauge transformation the Green's
function is not.
The Green's function for layer 1 is thus
multiplied by a phase factor
$\exp \left\{{ i\over \hbar} e
[\Lambda(\vec{r})-\Lambda(\vec{r^{'}})]\right\}$
giving
\begin{equation}
G_{2}^{+}(\vec{r},\vec{r^{'}})=
\exp\left \{{i \over \hbar} e \Lambda(\vec{r})\right\}
G_{1}^{+}(\vec{r},\vec{r^{'}})
\exp\left \{{-i \over\hbar}e \Lambda(\vec{r^{'}})\right\}
\hspace{3pt}. \label{gauge}
\end{equation}
Making use of this relationship we have
\begin{equation}
\sigma_{zz}={2 e^2 t_{c}^{2} c \over
\hbar \pi} \int{d^2r}
\hspace{4pt}\left| G_{1}^{+}(\vec{r},0,E_F)
\right|^2
\cos({{ec B \over \hbar}  \sin\theta y}) \ , \label{f1}
\end{equation}
where we have used the fact that
$\left| G_{1}^{+}(\vec{r},0,E_F)\right|^2 $
is translation invariant.

Note that Eq.~(\ref{f1}) is a very general
expression which holds
provided that intralayer momentum is conserved
and the interactions between the layers can be
neglected. It is valid in the presence of interactions
within the layers and for a non-Fermi liquid.\cite{clarke}
Second, this expression shows that for weakly incoherent
interlayer transport the interlayer conductivity
is completely determined by the {\it one}-electron
Green's function whereas the intralayer conductivity
is determined by {\it two}-electron Green's
functions\cite{macdonald}.

It is the averaging of the phase factor
over the spatial integral
in Eq.~(\ref{f1}) that gives rise
to the AMRO effect.
The length scale associated with the magnetic field
for the quasi-two-dimensional system is the
cyclotron length $R$ which at the Fermi energy is
$R = \hbar k_F / (e B \cos \theta)$.
For the quasi-one-dimensional case the length scale associated
with oscillations perpendicular to the chains
is $ R = 2 t_b/(e v_F B \cos \theta)$\cite{chaikin}.
At this length scale the phase difference
between the wave function
of adjacent layers is
$\Lambda(R) = e B \sin\theta c R = \gamma \tan \theta $.
Naively, we might expect maximum resistivity
when this phase difference
is an odd multiple of $\pi$, leading to a condition
different from Eq.~(\ref{minangle}).
However, one must take into account averaging
of the electron position over the perpendicular direction.

We now proceed to evaluate (\ref{f1}) for the simplest
possible situation, where there is a Fermi liquid
within each layer and the magnetic field is
small enough that we can take the semi-classical
limit of the Greens functions.

\subsection{Quasi-two-dimensional case}
The Green's function for layer 1, in the absence of
scattering can be written~\cite{kleinert}
\begin{eqnarray}
G(\vec{r},\vec{r^{'}},t)_1 & = & {m^*\over 2 \pi i \hbar t}
\hspace{3pt}
{\omega_{c} t/2 \over \sin(\omega_{c} t/2)}
\exp \left({i \omega_{c}\over \hbar } L \right)
\label{L1}
\end{eqnarray}
and
\begin{equation}
L = { m^*\over 2}\left[{|\vec{r}-\vec{r^{'}}
|^2 \over 2}\cot({\omega_{c} t \over 2})
+(x + x^{'})(y^{'}-y) \right] \ ,
\end{equation}
where $\omega_{c} = \omega_0 \cos\theta =
{e B \cos\theta / m^*}$ is the cyclotron frequency.
In order to calculate the conductivity we
follow the same approach that
Hackenbroich and von Oppen~\cite{hack}
used to study Shubnikov - de Haas oscillations
in two-dimensional electron systems.
In the presence of scattering the
energy-dependent Green's function is
\begin{equation}
G^{+}(\vec{r}, \vec{r^{'}}, E) =
{1 \over i \hbar} \int_{-\infty}^{\infty}
dt \exp\left({{i \over \hbar} (E + i\Gamma)t}\right)
\hspace{4pt} G(\vec{r},\vec{r^{'}},t) \hspace{3pt}. \label{G1}
\end{equation}
where $\Gamma = {\hbar \over 2 \tau}$ is the scattering rate.
The retarded Green's functon is obtained using
$G^{-}(\vec{r^{'}},\vec{r},E)=[G^{+}
(\vec{r},\vec{r^{'}},E)]^{*}$.
We perform the
integral in Eq.~(\ref{G1}) by the stationary phase
method which is valid in the
semi-classical limit($\hbar \rightarrow 0$).
The stationary phase condition gives
\begin{equation}
E = {m^* \omega_{c}^2 \over 8} \left
({|\vec{r} - \vec{r^{'}}|\over
\sin(\omega_{c} t/2)}\right)^2 \hspace{3pt}.
\end{equation}
This shows that if the cycloron radius
is $R_c$ then
$G^{+}(\vec{r} ,\vec{r^{'}}, E)$ vanishes
for $|\vec{r} - \vec{r^{'}}| > 2 R_c$,
while, for $|\vec{r} - \vec{r^{'}}| < 2 R_c$
there exists two different cyclotron
orbits and one finds an infinite set of
stationary times given by~\cite{hack}
\begin{equation}
T_{n,q}={2 \pi n \over\omega_{c}} + t_q \ ,
\end{equation}
where $n$ determines the number of
revolutions the electron makes to get
from $\vec{r}$ to $\vec{r^{'}}$ and
$q = S$ or $L$ denoting the two different
paths it can take (see Fig. {\ref{paths}). The
times to traverse these paths are calculated using the
the stationary phase condition and
$E= m^* R_{c}^{2} \omega_{c}^2 /2$, to give
\begin{eqnarray}
t_S & = & {2\over \omega_{c}}\arcsin\left
({|\vec{r} - \vec{r^{'}}| \over 2 R_c }
\right) \nonumber \\
\nonumber \\
t_L & = & {2\over \omega_{c}}\left[\pi -\arcsin\left
({|\vec{r} - \vec{r^{'}}|
\over 2 R_c }\right)\right] \hspace{3pt}. \label{TSL}
\end{eqnarray}
Putting all this together and performing the
integrals we obtain
\begin{eqnarray}
G_{1}^{+}(\vec{r}, \vec{r^{'}}, E) & = & { m^*\over 2 i \hbar}
\sum_{n=0}^{\infty} \sum_{q=S,L}
\left({\omega_{c} \over \pi i \hbar E
\sin(\omega_{c} T_{n,q})}\right)^{1\over 2}
\hspace{3pt}
\exp\left(-{T_{n,q} \over 2 \tau} \right)
\hspace{3pt} \nonumber \\
& & \times \exp\left({ {i\over \hbar}
S_{n,q} -{i \pi\over 2} \eta_{n,q} } \right) \ ,
\end{eqnarray}
where
\begin{equation}
S_{n,q}= E T_{n,q} +
{ m^* \omega_{c}\over 2}\left[{|\vec{r} - \vec{r^{'}}|^2
\over 2}\cot({\omega_{c} T_{n,q} \over 2})+
(x +x^{'})(y^{'}-y)\right]
\end{equation}
and $\eta$ is the Maslov or Morse index
(the number of conjugate points along the orbit
(Ref. \onlinecite{kleinert}, p.223)),
$\eta_{n,S}= 2n$ and $\eta_{n,L}=2n + 1$.
Equation~(\ref{f1}) can be written as
\begin{equation}
\sigma_{zz}={2 e^2 t_{c}^{2} c \over
\hbar \pi} \int{d^2|\vec{r}-\vec{r^{'}}|} \hspace{4pt}
\left| G_{1}^{+}(\vec{r},\vec{r^{'}},E_F)
\right|^2 \nonumber \\
\cos\left({ {ec \over \hbar} B\sin\theta
|\vec{r} - \vec{r^{'}}| \sin\phi}\right) \hspace{3pt} \ ,
\end{equation}
where $\phi$ is the angle between the vector
$|\vec{r} - \vec{r^{'}}|$ and the $x$-axis.
Substituting the semiclassical expressions
for the Green's functions into the above equation
and changing the integrals over $r$ and
$r^{'}$ to polar coordinates, one obtains a
double sum (denoted by the subscripts(1,2)
of $n$ and $q$) over the classical trajectories

\begin{eqnarray}
\sigma_{zz} & = & \left({m^* e t_{c} \over
\hbar^2 \pi}\right)^2
\left({ \omega_{c} c \over 2 E_F }\right)
\sum_{n_1, n_2 = 0}^{\infty} \hspace{2pt}
\sum_{q_1, q_2 = S, L}
\int_{0}^{2\pi} d\phi \int_{0}^{2R_c}
|\vec{r} - \vec{r^{'}}| d|\vec{r} - \vec{r^{'}}| \hspace{4pt}
\exp\left(-[T_{n_1, q_1}+T_{n_2, q_2}] \over
2\tau \right) \nonumber \\
& & \times {1 \over \sqrt{\sin(\omega_{c} T_{n_1, q_1})
\sin(\omega_{c} T_{n_2, q_2})}}
\exp\left({i\over \hbar}{[S_{n_2,q_2}-S_{n_1,q_1}]} -
{i \pi \over 2}{[\eta_{n_2,q_2}-\eta_{n_1,q_1}]} \right) \nonumber \\
& &\times \cos\left({ {e c B\over \hbar} \sin\theta
|\vec{r} - \vec{r^{'}}| \sin\phi}\right) \hspace{3pt}.
\end{eqnarray}
We make the simplification that $q_1$ = $q_2$,
for when $q_1 \neq q_2$ the integrand oscillates,
that is as $\hbar \rightarrow 0$ the
oscillations cancel each other
and therefore do not contribute. This gives
\begin{eqnarray}
\sigma_{zz} & = & \left({m^* e t_{c} \over
\hbar^2 \pi}\right)^2
\left({ \omega_{c} c \over 2 E_F }\right)
\int_{0}^{2\pi} d\phi \int_{0}^{2R_c}
|\vec{r} - \vec{r^{'}}| d|\vec{r} - \vec{r^{'}}|
\hspace{4pt}
\sum_{n_1, n_2 = 0}^{\infty}
\exp\left(-{\pi \over \omega_{c} \tau}{(n_1+n_2)}
\right) \nonumber \\
& &\times\left[ {\exp{(-t_S/\tau)}\over
\sin(\omega_{c} t_S)} +
{\exp{(-t_L/\tau)}\over \sin(\omega_{c} t_L)}
\right]
\cos\left( 2\pi \left[{E_F \over \hbar \omega_{c}}
- {1\over 2}\right]
(n_2-n_1) \right)
\cos \left({ {e c B \over \hbar}  \sin\theta
|\vec{r} - \vec{r^{'}}| \sin\phi}\right)\hspace{3pt}.
\end{eqnarray}
Terms with $n_1 \neq n_2$ correspond to the
Shubnikov-deHaas oscillations. We
neglect these by setting $n_1 = n_2$ since they
will be smaller that the leading order terms by a
factor of order $\exp({-{\pi\over\omega_c \tau }})$ and
thus we have
\begin{eqnarray}
\sigma_{zz} & = & A
\int_{0}^{2\pi} d\phi
\sum_{n=0}^{\infty} {\exp\left(-{2 \pi n\over
\omega_{c} \tau} \right)}
\left[ \int_{0}^{\pi / \omega_{c}}
\exp\left(-{t_S \over \tau}\right)
\hspace{3pt}
\cos\left(\eta \sin\left({\omega_{c} t_S \over
2}\right)\right)
\hspace{3pt} dt_S \right. \nonumber \\
& & + \left. \int_{\pi/\omega_{c}}^{2\pi/\omega_{c}}
\exp\left(-{t_L \over \tau}\right)
\hspace{3pt} \cos\left(\eta \sin\left({\omega_{c} t_L\over 2}
\right)\right) \hspace{3pt} dt_L
\right] \ ,
\end{eqnarray}
where $A ={ e^2 t_{c}^{2} m^* c \over \pi^2 \hbar^4}$
and $\eta =$
${2ec \over \hbar} B \sin\theta R_c \sin\phi=
2 c k_F \tan\theta \sin\phi =$
$2 \gamma \tan\theta \sin\phi$.
Combining the integrations over $t_S$ and
$t_L$ and performing the summation over $n$,
one obtains
\begin{equation}
\sigma_{zz} = {A \over (1-e^{-2 \pi / \omega_{c} \tau})}
\int_{0}^{2 \pi} d\phi
\left[ 
\int_{0}^{2\pi/\omega_{c}} \exp(-t/\tau) 
\cos\left(\eta \sin({\omega_{c} t\over 2})\right) 
dt \right] \hspace{3pt}. \end{equation} 
To evaluate the integral over $t$ we make use 
of the identity~\cite{abram}
\begin{equation}
\cos(\eta \sin\beta) = J_{0}(\eta) + 
2 \sum_{k=1}^{\infty} J_{2k}(\eta)\cos(2k\beta)\hspace{3pt}.
\end{equation}
The conductivity then simplifies to
\begin{equation}
\sigma_{zz}= A \tau \int_{0}^{2 \pi}
\left[ J_{0}(\eta) + 2 \sum_{k=1}^{\infty}
{J_{2k}(\eta)\over 1 +(k \tau \omega_{c})^2}
\right] d\phi \hspace{3pt}.
\end{equation}
This integral is of the form
$\int_{0}^{2\pi} J_{2k}(z \sin\phi) d\phi$,
where $z= 2 \gamma \tan\theta$, which can be
evaluated using
the relation~\cite{abram}
\begin{equation}
\int_{0}^{2 \pi} J_{2k}(z \sin\phi) d\phi =
2\pi J_{k}\left({ z / 2} \right)^{2}.
\end{equation}
We then
obtain an expression for the conductivity
which is of the form Eq.~(\ref{result}).
Previously, Yoshioka calculated the interlayer tunneling
of a quasi-two-dimensional system in the
absence of scattering.\cite{yosh}

\subsection{Quasi-one-dimensional case}
The Hamiltonian within a layer in a magnetic field is
\begin{equation}
H = \alpha i \hbar v_F {\partial \over \partial x} -
2 t_b \cos\left(b \left({1 \over i}{\partial \over
\partial y} - e x B \cos\theta \right)\right) \ ,
\end{equation}
where $\alpha =\pm 1$ denotes which sheet of the
Fermi surface the electron is on.
The wave function within a layer is given by\cite{yak}
\begin{equation}
\psi_{k_x, k_y, \alpha}(x,N,t) = \exp{ \left\{{i}
[-{\epsilon t \over \hbar}+ k_x x +b k_y N -
\alpha \lambda \sin(k_y b-qx)] \right\}} \ ,
\end{equation}
where $N$ denotes the number of the chain,
$x$ is the distance along the chain and
the dispersion relation
\begin{equation}
\epsilon_{\alpha}(k_x, k_y) = \alpha \hbar k_x v_F
\end{equation}
and
\begin{equation}
q={ e b B\cos\theta\over \hbar} = {\omega_B \over v_F}
\end{equation}
where $\omega_B$ is the oscillation frequency
given by (\ref{omb}) and
\begin{equation}
\lambda = {2 t_b \over e b v_F B\cos\theta} \hspace{3pt}.
\end{equation}
The transverse motion of the electrons due to the field is
approximately $\lambda b$.\cite{chaikin} The one-electron advanced
Green's function in
the absence of scattering is
\begin{equation}
G^{+}(x, x^{'}, N, N^{'}, t, 0) =
\sum_{k_x, k_y,\alpha} \psi_{k_x, k_y, \alpha}^{*}(x^{'},N,t)
\hspace{2pt} \psi_{k_x, k_y, \alpha}(x,N,0)
\end{equation}
for $t>0$. Taking the Fourier transform (with respect to time) of this and
including a scattering rate $\Gamma = {\hbar \over 2 \tau}$
\begin{equation}
G^{+}(x,x^{'},N, N^{'}, E) =
{1 \over i \hbar} \int_{0}^{\infty}
dt \exp \left\{ {i \over \hbar} \hspace{2pt}
(E + i \Gamma ) t \right\} \hspace{4pt} G^{+}(x, x^{'},
N, N^{'}, t, 0) \hspace{3pt}.
\end{equation}
After performing the integral in $t$ and $k_x$
\begin{equation}
G^{+} = -{i \over \hbar v_{F}}
\sum_{k_y, \alpha} { 1\over \alpha}
\exp \left\{ i [ b k_y (N-N^{'}) +
\alpha \lambda L ]\hspace{3pt} \right\}
\exp{\left\{{{i |x-x^{'}|\over \hbar v_{F}}}
\left( E + {i \hbar \over 2 \tau}\right)
\right\}} \ ,
\end{equation}
where
\begin{equation}
L = \sin(k_y b -q x^{'}) -\sin(k_y b - q x)\hspace{3pt}.
\end{equation}
This is similar to the quasi-classical Green's function
given by Gorkov and Lebed.\cite{gorkov}

The conductivity (\ref{condG}) then becomes
\begin{eqnarray}
\sigma_{zz} & = & { e^{2} t_{c}^{2} c \over
\hbar^3 v_{F}^{2} \pi L_x L_y}
\sum_{N, N^{'}} \sum_{k_{y1}, k_{y2}, \alpha}
\int dx \int dx^{'}
\left[ \exp \left\{ i (N-N^{'})
\left( k_{y1} b-k_{y2} b- {e b c B\sin\theta
\over \hbar}
\right) \right\} \right. \nonumber \\
& & \times \left. \exp \{i \alpha \lambda S_1 \}
\exp \left\{-{|x-x^{'}| \over v_{F} \tau} \right\} +
\hspace{10pt} c.c \hspace{10pt} \right] \ ,
\end{eqnarray}
where
\begin{equation}
S_1 = L_1 - L_2 =\sin(k_{y1} b-qx^{'})-\sin(k_{y1} b-qx)-
\sin(k_{y2} b-qx^{'})+\sin(k_{y2} b-qx)\hspace{3pt}.
\end{equation}
If we now let $M_{\pm}= N \pm N^{'}$ then the
sum over $M_{-}$ gives a delta function
$2 \pi \delta(k_{y1} b,k_{y2} b- {e b c B\sin\theta \over \hbar})$.
Replacing $k_{y1} b$ with $k_y b$
the conductivity then simplifies to
\begin{equation}
\sigma_{zz} = {e^{2} t_{c}^{2} c \over \hbar^3
\pi v_{F}^{2} L_x}
\sum_{k_{y}, \alpha} \int dx \int dx^{'}
\left[ \exp \{i \alpha \lambda S_2 \}
\exp \left\{-{|x-x^{'}| \over v_{F} \tau} \right\}
+ c.c \right] \ ,
\end{equation}
where
\begin{equation}
S_2 = \sin(k_{y}b-qx^{'})-\sin(k_{y}b-qx)-
\sin \left(k_{y}b- {e b c B\sin\theta \over \hbar}
-qx^{'} \right)
+\sin \left(k_{y}b-{e b c B\sin\theta \over \hbar}
- qx \right)\hspace{3pt}.
\end{equation}
Now shift $k_yb$ to $k_yb + qx^{'}$ and make the
substitution
$\Delta = {e b c B\sin\theta \over \hbar}$,
thus $S_2$ becomes
\begin{equation}
S_2 = \sin(k_{y}b)-\sin(k_{y}b-q(x-x^{'}))-\sin(k_{y}b-\Delta)
+\sin(k_{y}b-\Delta- q(x-x^{'}))\hspace{3pt}.
\end{equation}
We now let $x_{\pm} = {(x \pm x^{'}) \over 2}$
and $x_{-} = v_{F}t$ and perform the integral over $x_{+}$.
This simplifies the conductivity giving
\begin{equation}
\sigma_{zz} = {2 e^{2} t_{c}^{2} c \over \hbar^3 \pi v_{F} }
\sum_{k_{y},\alpha} \int_{0}^{\infty} dt
\exp \{i \alpha \lambda S_3 \}
\exp \left\{-{2 t \over \tau} \right\}
+\hspace{10pt} c.c \ ,
\end{equation}
where $S_3$ is given by

\begin{equation}
S_3 = \sin(k_{y}b)-\sin(k_{y}b-2\omega_B t)-\sin(k_{y}b-\Delta)+
\sin(k_{y}b-\Delta- 2 \omega_B t )
\end{equation}
and $\omega_B = q v_F$.
This can be separated into two parts and
simplified using the appropriate
trigonometric identities
\begin{eqnarray}
S_3 & \equiv & \mu - \beta \nonumber \\
& \equiv & 2 \cos \left( k_y b -
{\Delta \over 2}\right)
\sin \left({\Delta \over 2}\right)-
2 \cos \left( k_y b - {\Delta \over 2} -
2\omega_B t\right)
\sin \left({\Delta \over 2}\right) \nonumber \\
& = & \Delta \cos \left( k_y b -
{\Delta \over 2}\right) -
\Delta \cos \left( k_y b - {\Delta \over 2} -
2\omega_B t\right)
\end{eqnarray}
where we have taken $\Delta << 1$. We can
justify this by
considering the dimensions of the unit cell
and the magnetic flux passing
through the area of the cell. If the magnetic
field $ B \sim 10$ T, and
the area $bc \sim 10^{-18}$ m$^2$, then the
flux $\Phi = B bc $ will be small
and thus $\Delta = \Phi/ \Phi_0 << 1$ where
$\Phi_0 = \hbar / e$
is a flux quantum. Re-writing the
conductivity we obtain

\begin{equation}
\sigma_{zz} = {2 e^{2} t_{c}^{2} c
\over \hbar^3 \pi v_{F} }
\sum_{\alpha} \int_{0}^{\infty} dt
\exp \left\{-{2 t \over \tau} \right\}
\int_{-{\pi / b}}^{{\pi / b}} {dk_y \over 2 \pi}
\exp \left\{-i \alpha \lambda \beta \right\}
\exp \left\{i \alpha \lambda \mu \right\}
\hspace{3pt}+\hspace{3pt} c.c \hspace{3pt}.
\label{sigq1d_gen}
\end{equation}
Using the identity~\cite{abram}

\begin{equation}
\exp{\left[ {z\over 2}(h - {1 \over h}) \right]} =
\sum_{n = -\infty}^{\infty} h^{n} J_{n}(z)
\end{equation}
the exponentials in $\mu$ and $\beta$ become

\begin{eqnarray}
\exp \left\{-i \alpha \lambda \beta \right\} & = &
\sum_{n =-\infty}^{\infty} {-i^{n}}
J_n(\alpha \lambda \Delta )
\exp{\left\{i n\left[k_y b- {\Delta \over 2} -
2\omega_B t \right] \right\}} \nonumber \\
\exp \left\{i \alpha \lambda \mu \right\} & = &
\sum_{n^{'} =-\infty}^{\infty} {i^{n^{'}}}
J_{n^{'}}(\alpha \lambda \Delta )
\exp{\left\{i n^{'}\left[k_y b- {\Delta \over 2}
\right] \right\}}\hspace{3pt}.
\end{eqnarray}
Substitution of these into Eq.~(\ref{sigq1d_gen})
and performing
the integral in $t$ gives

\begin{eqnarray}
\sigma_{zz} & = & { e^{2} t_{c}^{2} \tau c
\over 2 \hbar^3 \pi^2 v_{F} } \sum_{\alpha}
\sum_{n =-\infty}^{\infty} \sum_{n^{'}=-\infty}^{\infty}
{-i^{n}} {i^{n^{'}}} J_n(\alpha \lambda \Delta )
J_{n^{'}}(\alpha \lambda \Delta) \nonumber \\
& & \times \left[{ 1 \over
{1 + in\omega_B \tau }}\right]
\int_{-{\pi / b}}^{{\pi / b}} dk_y
\exp{ \left\{ i \left(k_y b-{\Delta
\over 2}\right)[n^{'}+ n] \right\}}
\hspace{3pt}+\hspace{3pt} c.c \hspace{3pt}.
\end{eqnarray}
This integral is zero unless $n = -n^{'}$, thus

\begin{eqnarray}
\sigma_{zz} = { e^{2} t_{c}^{2} \tau c \over b
\hbar^3 \pi v_{F} } \sum_{\alpha}
\sum_{n =-\infty}^{\infty}
\left[{ J_n(\alpha z)^{2}
\over {1 + i n \omega_B \tau } }\right] + c.c
\end{eqnarray}
where $z = {\lambda \Delta \over \hbar} =
{2 c t_b \over \hbar v_F}{ B \sin\theta\over B\cos\theta}=
\gamma \tan\theta$ and $\gamma$ is the same          as for the
coherent case.
The summation over $\alpha$ is performed by noting
that $J_n(z)^2 = J_n(-z)^2$ for all $n$. Finally
we include the complex conjugate
to obtain
an expression which can be written as Eq.~(\ref{result}).

\subsection{The magnetic field parallel to the layers}

We consider here the field range over which the result
(\ref{linear}) holds for incoherent interlayer transport.
We define the magnetic field as $\vec{B}=(B_x,0, 0)$ and the
vector potential as $\vec{A}=(0, -zB_x,0)$.
It is easiest to work with spectral functions in the momentum
representation; the interlayer conductivity is then given by
\begin{equation}
\sigma_{zz} = {e^2 t_c^2 c\over \hbar \pi} \sum_k
A_1(\vec{k}, E_F) A_2(\vec{k}, E_F) \ ,
\end{equation}
where $A_1$ and $A_2$ are the spectral functions for the two layers.
Due to momentum conversation we can then write
the spectral function for layer two in terms of layer
one as
\begin{equation}
A_1(\vec{k}, E_F) = A_2 \left(\vec{k}-{e \over \hbar}
\vec{A}, E_F \right) = A(\vec{k}, E_F)\label{spec2}\ .
\end{equation}
where
\begin{equation}
A(\vec{k}, E_F)=  { 2 \Gamma \over
(E_F - \epsilon(\vec{k}))^2 + \Gamma^2} \label{spec1} \ ,
\end{equation}
and $\Gamma = \hbar / 2\tau$ and $\epsilon(\vec{k})$ is
the dispersion within the layer.

We now specialise to the quasi-two-dimensional case.
Substituting (\ref{spec1}) and (\ref{spec2})
into the conductivity gives
\begin{equation}
\sigma_{zz} = {e^2 t_c^2 c\over \hbar \pi^3}
\int dk_x dk_y
{\Gamma \over \left[E_F- {\hbar^2 \over2 m^*}
(k_x^2+k_y^2) \right]^2 +\Gamma^2}
{\Gamma \over \left[E_F- {\hbar^2 \over2 m^*}
\left(k_x^2 +\left(k_y +{ e\over\hbar }c B\right)^2
\right) \right]^2 +\Gamma^2} \label{cond_inlayer} \ .
\end{equation}
Now, introduce polar co-ordinates $(k,\phi)$ so
$k_x^2 + k_y^2 = k^2$,
$k_y = k \cos\phi$, and define
 $\Delta \equiv (ecB / \hbar)^2$ so 
 that Eq.~(\ref{cond_inlayer}) becomes
\begin{equation}
\sigma_{zz} = {e^2 t_c^2 c \over\hbar \pi^3}
\int_{0}^{\infty} k dk \int_{0}^{2 \pi} d\phi
{\Gamma \over \left[{\hbar^2 \over 2 m^*}k_F^2-
{\hbar^2 \over2 m^*}k^2 \right]^2 +\Gamma^2}
{\Gamma \over \left[{\hbar^2 \over2 m^*}k_F^2-
 {\hbar^2 \over2 m^*}
\left(k^2 + \Delta +{ 2 e\over\hbar }c Bk\cos\phi\right)
\right]^2 +\Gamma^2} \label{cond_inlayer2}
\end{equation}
Suppose that the field is sufficiently large that
$\Gamma \ll \hbar eB k_F c/m^* $
(which corresponds to $\omega_0 \tau \gamma \gg 1$)
then the first spectral function has a sharp peak 
near $k=k_F$  whereas near that peak the
second term varies slowly.  Hence, we set $k=k_F$ in the
second term and then integrate over $k$
to give
\begin{equation}
\sigma_{zz} = {e^2 t_c^2 c \Gamma m^* \over \pi^2 \hbar^3 }
 \int_{0}^{2 \pi} d\phi
{1 \over \left[{\hbar^2 \over2 m^*}
\left(\Delta + { 2 e\over\hbar }c B k_F \cos\phi\right)
 \right]^2 +\Gamma^2} \ .
\end{equation}
When $\Delta \ll ecB k_F / \hbar$, this integral will be
dominated by the behaviour near the two zeros of $\cos\phi$
so we can write the integral as
\begin{equation}
\sigma_{zz} = { 2 e^2 t_c^2 c \Gamma m^* \over \pi^2 \hbar^2  }
 \int_{-\infty}^{\infty} d\phi
{1 \over \left[{ \hbar \over m^* }e c Bk_F\phi
 \right]^2 +\Gamma^2}.
\end{equation}
Performing the integral gives (\ref{linear}) resulting
in a magnetoresistance which is linear in field.

When $\Delta \sim ec B k_F / \hbar$ that is $ecB / \hbar k_F \sim 1$
deviations from this linear in field behaviour will occur.
If $c\sim10 \AA$, $c k_F \sim 3$, then $B \approx 2000$ T.
Similar arguments apply to the quasi-one-dimensional case.
It will be shown in the next section that for coherent
interlayer transport the deviations from linear dependence
can occur at much lower fields.

\section{DEFINITIVE TESTS FOR COHERENT INTERLAYER TRANSPORT}
\label{sig}

We have shown that the Yamaji and Danner oscillations
exist for both coherent and weakly incoherent interlayer
transport and so cannot be used to establish
that the Fermi surface is three-dimensional.
We now consider three properties which are
different for coherent and incoherent interlayer transport.

\subsection{Beats in magnetic oscillations}
For quasi-two-dimensional systems
definitive evidence for the existence of a
three-dimensional Fermi surface,                    
is the observation of
a beat frequency in de Haas-van Alphen
and Shubnikov - de Haas oscillations.\cite{mckenzie}
The frequency $F$ of these oscillations is determined
by extremal areas $A$ of the Fermi surface,
$F = \hbar A /(2 \pi e) (Ref. $\onlinecite{wos}).
For the warped cylindrical Fermi surface 
(see Fig. 1 in Ref. \onlinecite{mckenzie})
there are two extremal areas, corresponding to
the ``neck'' and ``belly'' orbits. The small difference between
the two areas leads to a beating of the corresponding
frequencies $F_1$ and $F_2$. In a tilted
magnetic field the frequency difference is
\begin{equation}
{F_1 - F_2 \over F_1} = {4 t_c \over E_F}
J_0(k_F c \tan \theta)\ .
\label{beat}
\end{equation}
Table \ref{table2} lists several materials in which
such beat frequencies have
been observed.
In $\beta$-(BEDT-TTF)$_2$I$_3$ and
$\beta$-(BEDT-TTF)$_2$IBr$_2$ the angular
dependence of the beat frequency is consistent with (\ref{beat})
and $t_c/E_F \simeq 1/175$ and 1/280, respectively\cite{wosnitza6}.

However, Table \ref{table2} indicates
that in many other quasi-two-dimensional
organic metals no beat frequency has been observed.
This could be because the interlayer transport is incoherent
or because the interlayer hopping $t_c$ is
so small that the beats cannot be resolved
experimentally. Suppose that oscillations
but no beats are seen in the field range from $B_{min}$ to $B_{max}$.
This means that $\cos(2\pi (F_1 - F_2)/B)$ has no
zeroes in this field range, implying that
\begin{equation}
F_1 - F_2 < {B_{min} B_{max} \over B_{max} - B_{min}}\ .
\label{criterion}
\end{equation}
This together with Eq.~(\ref{beat}) can be used to establish an upper
bound for $t_c/E_F$.
For $\kappa$-(BEDT-TTF)$_2$I$_3$ the absence of beating has been used
to establish $t_c/E_F < 1/3000$\cite{wos,balthes}.
This implies a resistivity anisotropy
$\rho_{xx}/\rho_{zz} \sim (t_c/E_F)^2 < 10^{-7}$.
However, the observed\cite{dressel} anisotropy in the
$\kappa$-(BEDT-TTF)$_2$X materials is about
$10^{-3}$.
This inconsistency suggests that the interlayer transport
may be incoherent in $\kappa$-(BEDT-TTF)$_2$I$_3$.
However, it could be that the measured value of
$10^{-3}$ is too large because resistivity anisotropy
is too large because the measurement of $\rho_{xx}$
involves some component of $\rho_{zz}$
due to an imhomogeneous current distribution
or the current path being changed by sample defects.

\subsection{Peak in the angle-dependent magnetoresistance at 90 degrees}

Numerical solutions of Chambers' formula (\ref{chambers2}) for
coherent interlayer transport show that
for both quasi-one-dimensional\cite{dan} and
quasi-two-dimensional\cite{hanasaki} materials,
at sufficiently high fields, the angle-dependent magnetoresistance
has a peak as the field direction approaches
the layers (i.e., at $\theta=$ 90 degrees).
This peak is absent for incoherent interlayer transport
(see Figs. \ref{plots1} and \ref{plots2}).
Hanasaki et al.\cite{hanasaki} identified the peak
as being due
to closed orbits which occur when the field is parallel to the layers.
These orbits are associated with the cyclotron frequency
\begin{equation}
\Omega = \omega_0 \left({ 2 t_c m c^2 \over \hbar^2}\right)^{1/2}
= \omega_0 \gamma \left({ t_c \over E_F} \right)^{1/2}
\label{freq}
\end{equation}
and so will only be important when the field is
sufficiently large that $\Omega \tau > 1$.

Table \ref{table2} lists whether or not the peak has been
observed for a range of quasi-two-dimensional metals.
Note that the presence (absence) of the peak is not always
consistent with the observed presence (absence) of
beats. This can be because
the two sets of measurements were done on
different samples of different purity
(and thus had different values of $\tau$) or because the field
was not large enough to observe the peak.
The presence of a peak at 90 degrees in the AMRO data\cite{dan}
for (TMTSF)$_2$ClO$_4$ suggests that it has coherent
interlayer transport.

\subsection{Crossover from linear to quadratic field
dependence for a magnetic field parallel to the layer}

Schofield, Wheatley and Cooper\cite{scho} considered the
interlayer magnetoresistance for quasi-two-dimensional
systems with coherent interlayer transport and a
magnetic field parallel to the layer.
Eq. (25) of Reference \onlinecite{scho} gives an
expression for the interlayer conductivity
for all values of the magnetic field.
They showed that
when $\Omega \tau \ll 1$ the magnetoresistance
increases linearly with field, as in Eq. (\ref{linear}).
However, for $\Omega \tau \gg 1$
the field dependence becomes quadratic and is given by
\begin{equation}
{\sigma_{zz}(B) \over \sigma_{zz}(0)}
= { 1.96 \over (\gamma \omega_0 \tau)^2 }
\left({E_F \over t_c} \right)^{1/2}\ .
\end{equation}
The deviations from linear behavior will occur when
$\Omega \tau > 1$, i.e, $\omega_0 \tau > { 1 \over \gamma} 
\left({E_F \over t_c} \right)^{1/2}$. For typical
organic samples this will happen in the field range of
10-100 tesla.
In contrast for incoherent interlayer transport 
it was shown in Section IV C that
the deviation from the linear field dependence
would not occur until about 2000 T.
We are unaware of any material in which a search for this
linear to quadratic crossover has been made.
This field dependence is to be contrasted to that
at angles slightly different from 90 degrees,
which will be given by Eq. (\ref{linear}).
The ratio of these two expressions provides a means
to determine $t_c/E_F$ since $\gamma$ and $\omega_0 \tau$
can be deduced from AMRO data.

\section{CONCLUSIONS}

We have presented detailed calculations of
the interlayer magnetoresistance of quasi-one- and
quasi-two-dimensional Fermi liquids in a tilted magnetic
field. Two distinct models were used for the
interlayer transport. The first involved coherent
interlayer transport and made use of results
of semi-classical or Bloch-Boltzmann transport
theory. The second model involved weakly incoherent
interlayer transport where the electron is scattered many
times within a layer before coherently tunneling into
the next layer.
We found that the dependence of the interlayer magneoresistance
on the direction of the magnetic field is identical
for both models except when the field is almost parallel
to the layers.
An important implication of this result is that coherent
transport is not necessary for the observation
of the Yamaji and Danner oscillations.
Hence, observation of one of these effects in a particular
material cannot be interpreted as evidence that the material has a
three-dimensional Fermi surface.
Instead, we propose three unambiguous tests for
coherent interlayer transport:
(i) a beat frequency in the magnetic oscillations in
quasi-two-dimensional systems,
(ii) a peak in the angular-dependent magnetoresistance
when the field is parallel to the layers,
and (iii) a crossover from a linear to a quadratic field
dependence for the interlayer magnetoresistance when
the field is parallel to the layers.
A survey of published experimental data on a wide range
of quasi-two-dimensional organic metals suggests that
some have properties (i) and (ii) and others
do not.

In future publications we will examine the frequency dependent
interlayer conductivity and the Lebed and third angular effects
in quasi-one-dimensional systems.
A much greater challenge
is to explain the AMRO observed
in (TMTSF)$_2$PF$_6$ at pressures of about 10 kbar\cite{dan2,naughton}
and in the low-temperature phase of
$\alpha$-(BEDT-TTF)$_2$MHg(SCN)$_4$[M=K,Rb,Tl].
The angular dependence of the latter is inverted
compared to that of the Yamaji effect.
In particular, the magnetoresistance is smallest
when the field is in the layers,
the opposite of what one expects based on
the simple Lorentz force arguments relevant
to semi-classical magnetoresistance.
Understanding this may require
knowledge of the effect of an orbital magnetic field
on a strongly correlated electron system.
Little is known about this problem
except in the extreme quantum limit
of a partially filled lowest Landau level\cite{renn} which is
far from the situation considered here where usually
of the order of tens of Landau levels are filled.

\acknowledgements
This work was supported by the Australian Research Council,
the Australian Department of Industry, Science and Technology,
and the USA National High Magnetic Field Laboratory
which is supported by NSF Cooperative Agreement
No. DMR-9016241 and the state of Florida.
We thank C. C. Agosta,
P. W. Anderson, N. Bonesteel, J. S. Brooks, P. M. Chaikin,
E. I. Chashechkina, S. Hill, B. E. Kane, I. J. Lee, A. H. MacDonald, J.
Merino,
K. A. Moler, and J. S. Qualls
for helpful discussions.
We thank J. Wosnitza for helpful comments on the manuscript
and providing experimental data prior to publication.

\vskip - 0.4 truecm


\begin{figure}
\centerline{\epsfxsize=9cm \epsfbox{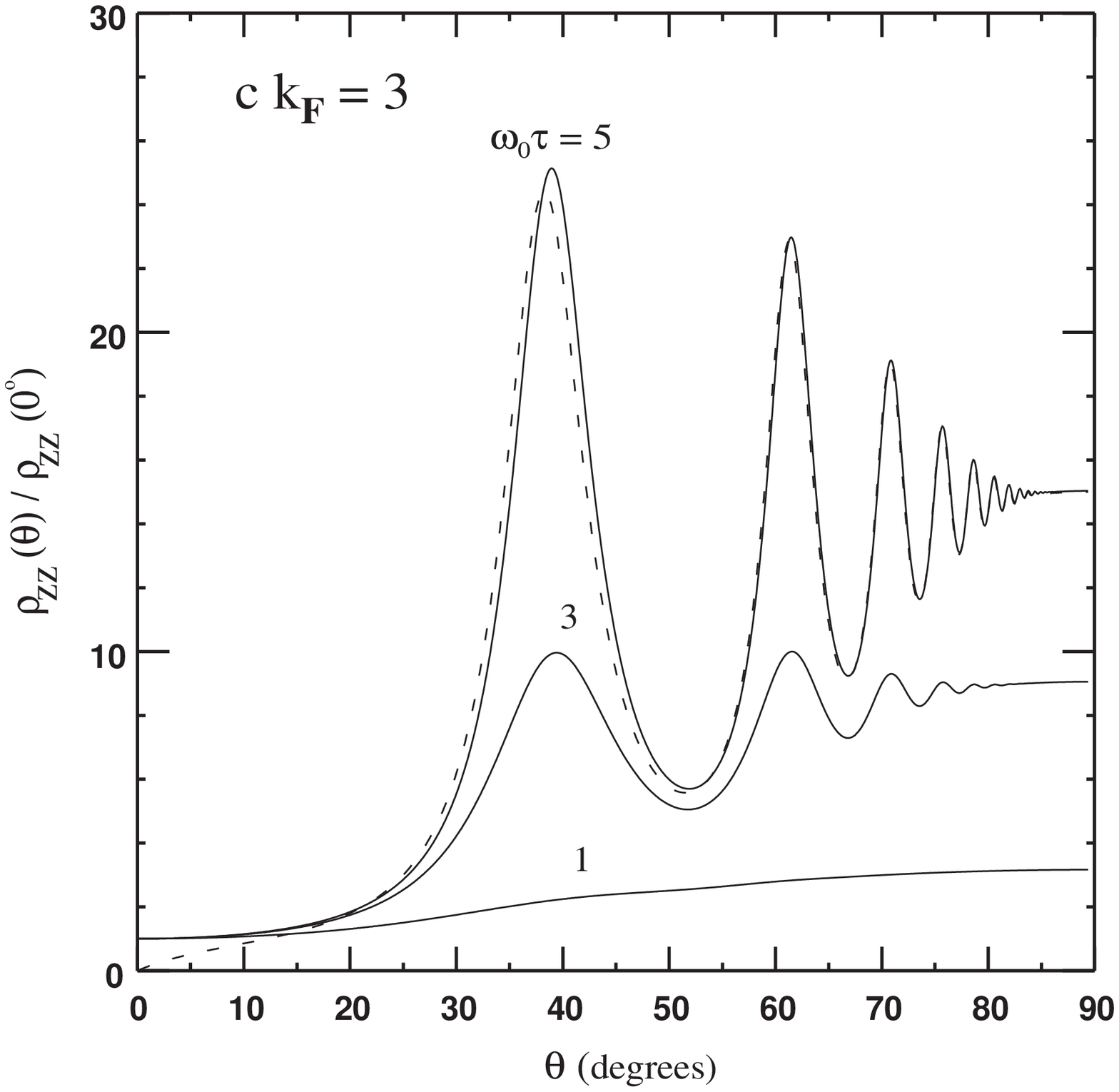}}
\caption{ 
Angular-dependent magnetoresistance oscillations.
The dependence of the interlayer resistance for a
typical quasi-two-dimensional system on
the direction of the magnetic field is shown for a range
of magnetic fields.
$\theta$ is the angle between the field and
the normal to the layers.
$\tau$ is the scattering time
within the layers and $\omega_0$ is the cyclotron fr
equency when
the field is perpendicular to the layers.
The curves shown are plots of Eq. (1) which
is valid for all $\theta$ for incoherent interlayer
transport and for all $\theta$ except close to 90 degrees
for coherent interlayer transport.
Note that the location of the maxima and minima is
independent of the field and the scattering time.
The dashed curve is a plot of the asymptotic expression
(\protect\ref{bs12}),
which can be seen to be a very good approximation
for $\theta > 20 $ degrees.
\label{plots1}}
\end{figure}

\begin{figure}
\centerline{\epsfxsize=9cm \epsfbox{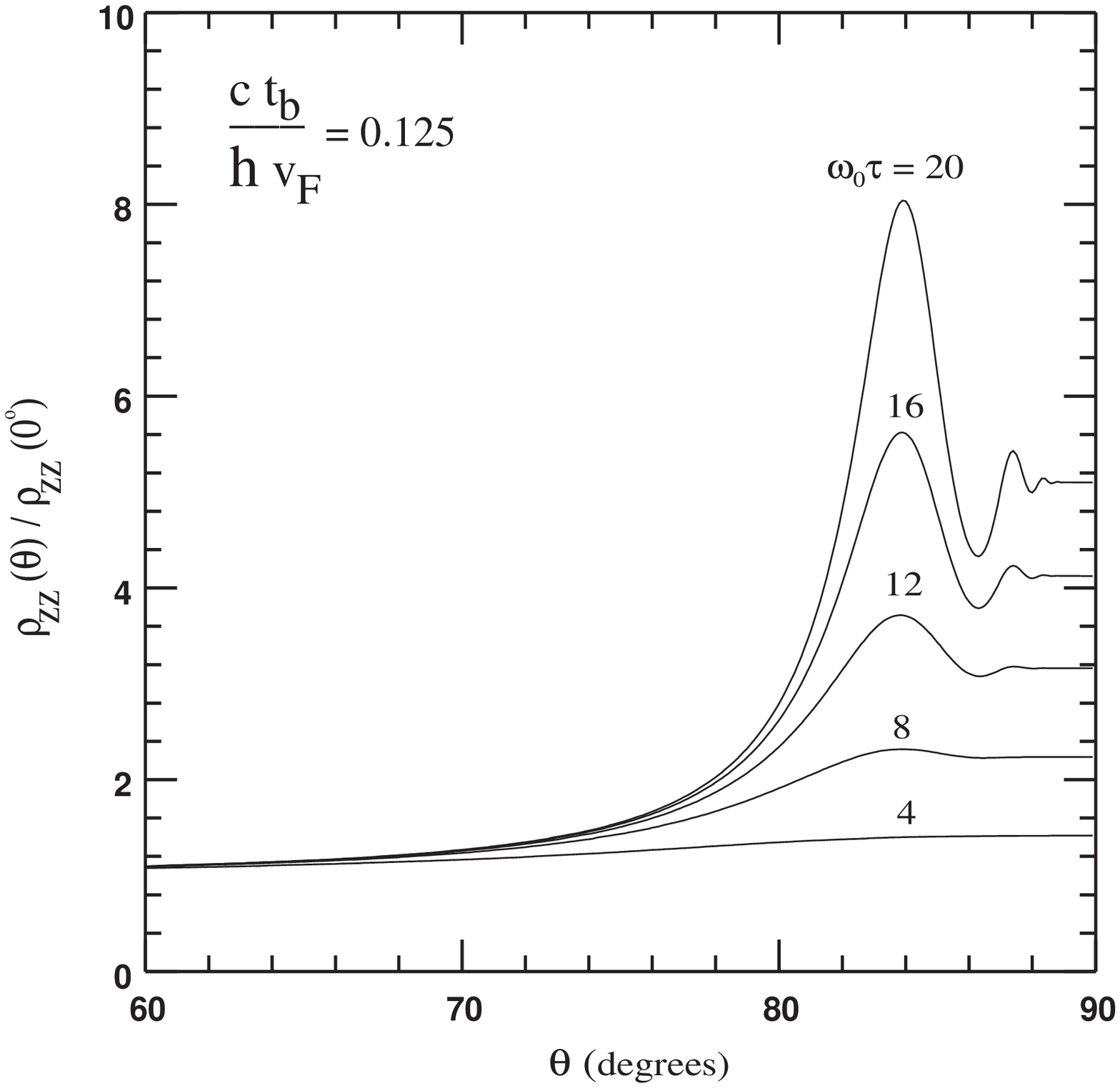}}
\caption{ 
Dependence of the interlayer resistance of
a quasi-one-dimensional system on
the direction of the magnetic field.
$\theta$ is the
 angle between the magnetic field and
the least conducting direction, with the field in the
same plane as the most conducting direction.
The parameter which defines the anisotropy of
the intralayer hopping $\gamma \equiv { 2 c t_c \over \hbar v_F }=0.25$.
$\tau$ is the intralayer scattering time
and $\omega_0$ is the frequency at which the electrons
oscillate between the chains when the field is
perpendicular to the layers.
Except very close to 90 degrees
this figure is similar to the experimental
data on (TMTSF)$_2$ClO$_4$ in Ref. \protect\onlinecite{dan}.
\label{plots2}} \end{figure}

\begin{figure}
\centerline{\epsfxsize=9.3cm \epsfbox{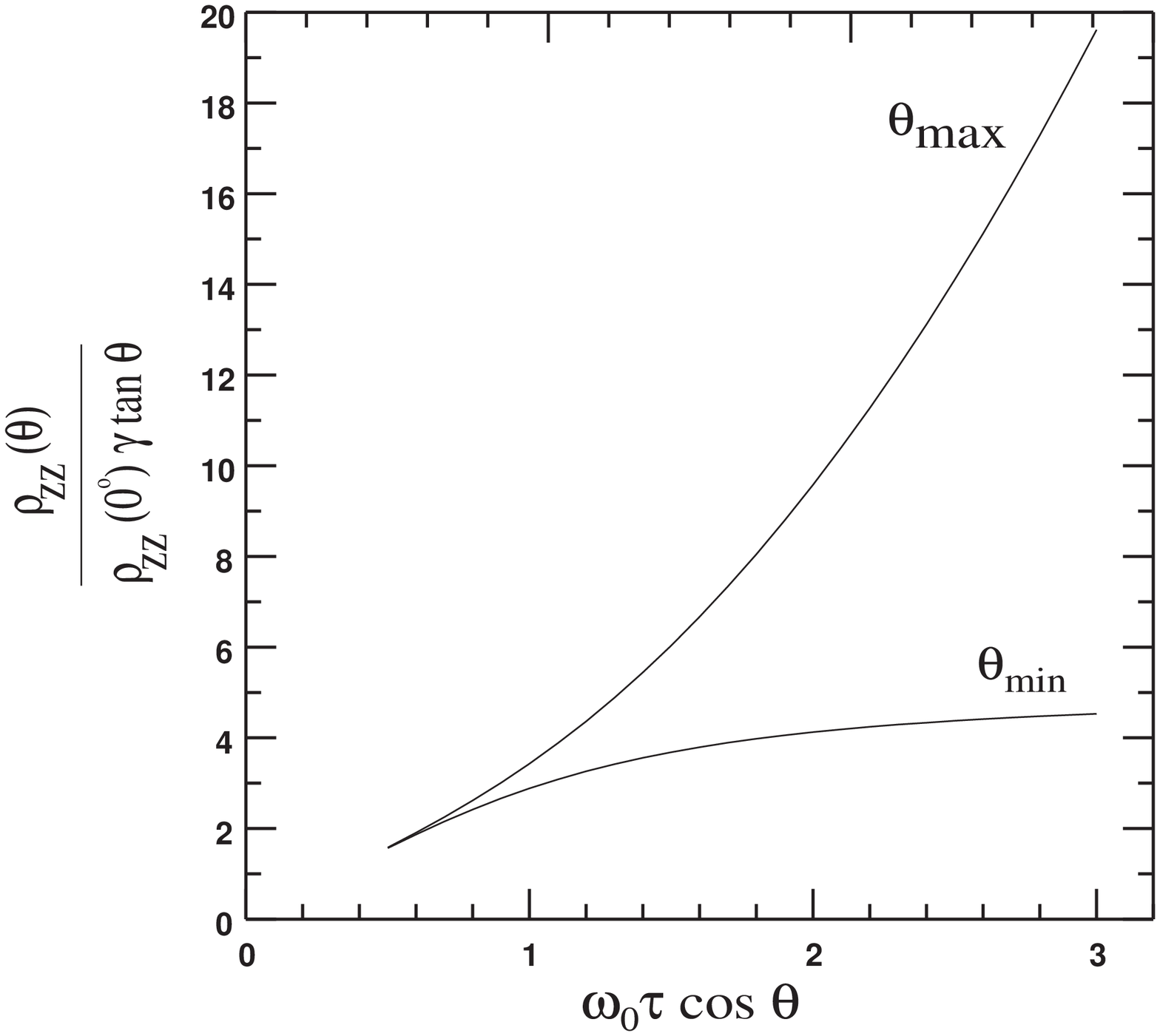}}
\caption{
Universal dependence of the interlayer resistivity on
the magnetic field and scattering time when the
field is tilted at an angle corresponding
to an AMRO minimum ($\theta_{min}$) and
an AMRO maximum ($\theta_{max}$).
For high fields the resistivity at the minima
becomes independent of field and has
the same temperature dependence as the
zero-field resistivity.
For high fields the resistivity at the maxima
increases quadratically with field and has
the same temperature dependence as the
inverse of the zero-field resistivity.
The curves are not plotted for small $\omega_0 \tau \cos \theta$
because the results derived in the
text are not valid in
that regime.
\label{p2theta}}
\end{figure}

\begin{figure}
\centerline{\epsfxsize=9.3cm \epsfbox{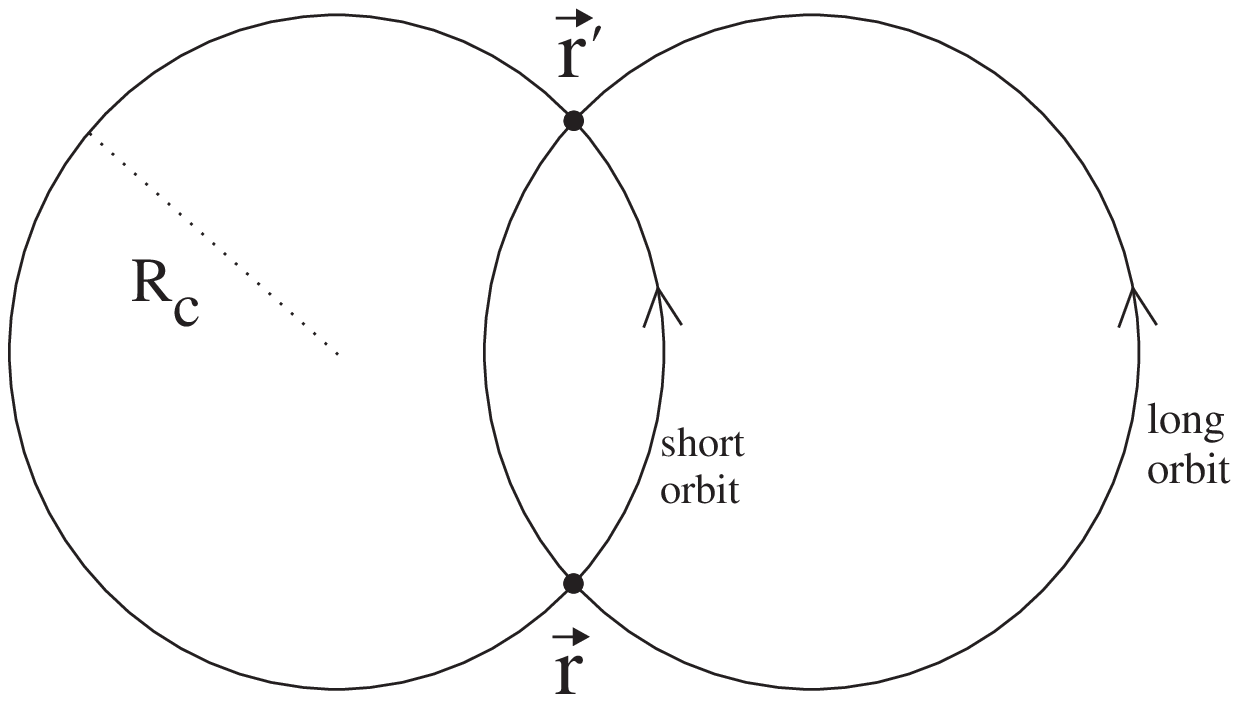}}
\vskip 0.5cm
\caption{Short and long semi-classical orbits joining
two points within a layer of a quasi-two-dimensional system
in a magnetic field perpendicular to the layers.
$R_c$ is the radius of the cyclotron orbit.
\label{paths}}
\end{figure}

\newpage
\onecolumn
\widetext

\begin{table}
\caption{
For a range of quasi-two-dimensional materials we
list whether or not beats in magnetic oscillations and a peak in
the angular dependent magnetoresistance at 90 degrees
has been observed.
For coherent interlayer transport both these features
should be present provided a wide enough range of magnetic
fields is explored.
A question mark indicates that the measurement has not
been made.}
\begin{tabular}{lll}
& Beats & Peak at 90 degrees \\
\tableline
$\alpha$-(BEDT-TTF)$_2$NH$_4$Hg(SCN)$_4$ &
no\protect\cite{osada} & no \protect\cite{osada} \\
$\alpha$-(BEDT-TTF)$_2$KHg(SeCN)$_4$ &
no\protect\cite{kovalev} & no\protect\cite{kovalev} \\
$\alpha$-(BEDT-TTF)$_2$KHg(SCN)$_4$ above 20 T &
no\protect\cite{harr}
& no\protect\cite{house} \\
$\alpha$-(BEDT-TTF)$_2$TlHg(SeCN)$_4$ &
no\protect\cite{goll,laukhin} & ? \\
$\alpha$-Et$_2$Me$_2$N[Ni(dmit)$_2$]$_2$ & ?
& yes\protect\cite{tajima} \\
$\alpha$-(BEDT-TSF)$_2$KHg(SCN)$_4$ above 6 kbar &
yes\protect\cite{agosta} & ? \\
$\beta_H$-(BEDT-TTF)$_2$I$_3$
& yes\protect\cite{kang,wosnitza5} & yes\protect\cite{hanasaki}\\
$\beta$-(BEDT-TTF)$_2$IBr$_2$
& yes\protect\cite{wosnitza6,kart1} & yes\protect\cite{kart1}\\
$\kappa$-(BEDT-TTF)$_2$I$_3$
& no\cite{schweitzer} & yes\protect\cite{helm} \\
$\kappa$-(BEDT-TTF)$_2$Cu$_2$(CN)$_3$ at 7 kbar
& ? & yes\protect\cite{ohmichi} \\
$\kappa$-(BEDT-TTF)$_2$Cu(SCN)$_2$
& no\protect\cite{wosnitza2} & ?\protect\cite{hc2}\\
$\theta$-(BEDT-TTF)$_2$I$_3$
& no\cite{terashima} & yes\protect\cite{kajita} \\
Sr$_2$RuO$_4$ & yes\protect\cite{mack} & yes\protect\cite{ohmichi2}
\end{tabular}
\label{table2}
\end{table}


\begin{references}
\bibitem{caveat}
We neglect terms involving non-diagonal components
of the resistivity tensor. Their contribution
to $\rho_{zz}$ can be neglected for weakly coupled layers.
\end{references}
\end{document}